%% file: labs2.tex
\documentclass[a4paper,twocolumn]{article}

\title{On the ground states of the Bernasconi model}

\author{Stephan~Mertens$^1$ and Christine~Bessenrodt$^2$\\[1ex]
 \small\it $^1$Institut f.\ Theoretische Physik, Otto-von-Guericke Universit\"at, Postfach 4120, 
  D-39016 Magdeburg, Germany\\
\small\it $^2$Institut f.\ Algebra u.\ Geometrie, Otto-von-Guericke Universit\"at, Postfach 4120, 
  D-39016 Magdeburg, Germany
}
                         
\usepackage{cite,graphicx,fancyheadings,float,theorem}
\usepackage[centertags]{amstex}

\newcommand{\myref}[1]{(\ref{#1})}

\theoremstyle{plain} 
\newtheorem{Fact}{Fact}
\newtheorem{Con}{Conjecture}

\date{\today}

\floatstyle{ruled}
\restylefloat{table}
\restylefloat{figure}

\begin{document}

\twocolumn[

\maketitle

\begin{center}
\begin{minipage}{14cm}
{\it
The ground states of the Bernasconi model are binary $\pm1$ sequences
 of length $N$ with low autocorrelations. We introduce the notion of perfect sequences,
binary sequences with one-valued 
off-peak correlations of minimum amount. If they exist, they are ground states. Using
results from the mathematical theory of cyclic difference sets, we specify all values of $N$ 
for which perfect sequences do exist and how to construct them. For other values of $N$,
we investigate almost perfect sequences, i.e.\ sequences with
two-valued off-peak correlations of minimum amount. Numerical and analytical results 
support the conjecture that almost perfect sequences do exist 
for all values of $N$, but that they are not always ground states.
We present a construction for low-energy configurations that works if $N$ is
the product of two odd primes.
\hspace*{2cm}
}
\end{minipage}
\end{center}
]

\input{intro}

\input{differences}
\input{perfect}

\input{almost}
\input{conclusions}

\bibliographystyle{unsrt}

\bibliography{strings,math,labs}

\end{document}

%% file: intro.tex
\section{Introduction}
\label{sec:intro}

Binary sequences of $+1$ and $-1$ with low autocorrelations have many applications
in communication engineering \cite{schroeder:84}. Their construction has a long 
tradition \cite{golay:82}, and has turned out to be a very hard
mathematical problem. Bernasconi \cite{bernasconi:87} introduced an Ising spin model 
that allows to formulate the construction problem in the framework of statistical
mechanics.

Consider a sequence of binary variables or Ising spins of length $N$,
\begin{equation}
  \label{eq:def_S}
  S = (s_0,s_1,\ldots,s_{N-1})\qquad s_i=\pm 1,
\end{equation}
and their autocorrelations
\begin{equation}
  \label{eq:periodic_C}
  C_g(S) = \sum_{i=0}^{N-1} s_is_{i+g},
\end{equation}
where all indices are taken modulo $N$. Bernasconi defined a Hamiltonian $H(S)$ 
by
\begin{eqnarray}
  \label{eq:hamiltonian}
  H(S) &=& \sum_{g=1}^{N-1} C_g^2(S) \\\nonumber
       &=& \sum_{i,j=0}^{N-1}\sum_{g=1}^{N-1}s_is_{i+g}s_js_{j+g}.
\end{eqnarray}
The ground states of this model with its long-range, four-spin interactions are 
the low autocorrelation binary sequences we are looking for. 

The Bernasconi model is completely deterministic in the sense that there is no
explicit or quenched disorder like in spin-glasses. Nevertheless
the ground states are by definition highly disordered. This
self-induced disorder resembles the situation in real glasses. In fact,
the Bernasconi model exhibits features of a glass transition like a
jump in the specific heat \cite{bernasconi:87}, and slow dynamics and aging
\cite{krauth:mezard:95}.

The replica scheme, an analytical method well established in spin-glass
theory, is of little use for deterministic systems.
People have approximated the Bernasconi model
by a model with explicit, quenched disorder,
which in turn can be analyzed analytically with the replica method 
\cite{bouchaud:mezard:94,marinari:parisi:ritort:94a}. This approach is valid
only in the high temperature regime, however, and provides only estimates for
the ground state energies.

Bernasconi introduced his model in order to apply statistical mechanics, especially simulated 
annealing,
to find low autocorrelation binary sequences. His approach turned out to be of little 
practical use, however: the energy minima found with the annealing procedure differ by factor 2 from
the conjectured ground state energy. This failure is due to the peculiar ``golf course''
property of the energy landscape which is dominated by a large number of local minima, while the 
global minima are deep and shallow holes, extremely isolated in configurations space.

The original Bernasconi model is defined with aperiodic autocorrelations, i.e.\ with
$N-1-g$ as the upper summation limit in equation \myref{eq:periodic_C}.
For this model, exhaustive enumeration of all $2^N$ configurations seems to be the only means
to get true ground states \cite{mertens:96b,labs:data:open}.

For the Bernasconi model with periodic autocorrelations, the situation is better. 
The tight connection between the
correlations of periodic binary sequences and mathematical objects called
{\em cyclic difference sets} can be exploited to get some exact results on the ground states.
The theory of difference sets is well established in mathematics but hardly known in 
statistical mechanics. It is one objective of this paper to fill this gap.
It will turn out, that ground states of the Bernasconi model can be constructed
from cyclic difference sets for certain values of $N$. Other values of $N$ require
some generalizations. Guided by numerical results, we will discuss such generalizations.

The paper is organized as follows: We start with a derivation of the equivalence
of autocorrelations in binary sequences and cyclic difference sets in section 
\ref{sec:difference_sets}.
In section \ref{sec:perfect_sequences}, we introduce {\em perfect sequences}, which are
ground states of the Bernasconi model -- if they exist. The existence problem in turn can be answered
using results from the theory of difference sets.
Numerical and analytical investigations of those cases, which are not
covered by perfect sequences, can be found in section \ref{sec:beyond_perfect_sequences}.
Section \ref{sec:conclusions} comprises our conclusions.

%% file: differences.tex
\section{Difference sets}
\label{sec:difference_sets}

Let $G = \Bbb{Z}_N$, the {\em cyclic group} of integers $0,\ldots,N-1$ with addition modulo $N$, and
let $D\subseteq G$ be a $k$-element subset of G. Each element $g\in G$ has a number of
different representations $g = d_1-d_2$ with $d_1,d_2\in D$, denoted as 
{\em replication number} $\lambda(g)$:
\begin{equation}
  \label{eq:def_lambda}
  \lambda(g) = \big| \{(d_1,d_2)\, |\, g = d_1-d_2,\,\, d_1,d_2\in D\}\big|.
\end{equation}
The trivial case is $\lambda(0) = k$. The values $\lambda(g)$ for
$g\neq0$ depend on $D$. The total number of differences $d_1-d_2$ equals $k^2$, leading
to the constraint
\begin{equation}
  \label{eq:constraint}
  k^2 = \sum_{g\in G}\lambda(g).
\end{equation}
Now consider the periodic binary sequence $(s_i)$ associated with $D$:
\begin{equation}
  \label{eq:SfromD}
  s_i = \left\{
    \begin{array}{ll}
      +1 & i\bmod N\in D\\
      -1 & i\bmod N\not\in D
    \end{array}
  \right..
\end{equation}
A straightforward calculation shows that
the autocorrelations of this sequence are given by
\begin{equation}
  \label{eq:C_N_k_lambda}
  C_g = N - 4(k-\lambda(g)).
\end{equation}
The converse is also true: Given a $\pm 1$ sequence of length $N$ with periodic 
autocorrelations $C(g)$,
each $g\in G$ has $\lambda(g) = k + (C(g)-N)/4$ representations $g=d_1-d_2$ within
the $k$-element set
\begin{equation}
  \label{eq:DfromS}
  D := \{i : s_i = +1, i = 0,\ldots,N-1\},
\end{equation}
so we conclude that
the periodic autocorrelations of a binary sequence and the number of difference representations 
of elements of a cyclic group $G$ in $D\times D$ are equivalent.

The definition of a replication number and its correspondence to the correlations of
periodic binary sequences do not depend on the 
concrete realisation of our group $G$ as 
$G=\Bbb{Z}_N$.
Let $G$ be a 
%
cyclic group of order $N$, written in multiplicative notation.
The replication number $\lambda(g)$ is defined as
\begin{equation}
  \label{eq:def_lambda_mult}
  \lambda(g) = \big| \{(d_1,d_2)\, |\, g = d_1\cdot d_2^{-1},\, d_1,d_2\in D\}\big|
\end{equation}
for a $k$-element subset $D\subseteq G$. To construct a periodic binary sequence 
based on $D$, we choose a 
generator 
$\alpha$ of the cyclic group $G$
and define the binary sequence to be
\begin{equation}
  \label{eq:SfromD_mult}
  s_i = \left\{
    \begin{array}{ll}
      +1 & \alpha^i\in D\\
      -1 & \alpha^i\not\in D
    \end{array}
  \right..
\end{equation}
Again
the correlations of this sequence are given by eq.~\myref{eq:C_N_k_lambda}.

If the replication number 
%
is constant for $g\neq0$, 
i.e.
\begin{equation}
  \label{eq:def_difference_set}
  \lambda(g) = \left\{
    \begin{array}{ll}
      k & \textrm{ for } g = 0\\
      \lambda & \textrm{ for } g \neq 0
    \end{array}
  \right.,
\end{equation}
the set $D$ is called an $(N,k,\lambda)$ {\em cyclic difference set}, where the term 
cyclic refers to the fact that the
underlying group $G$ is cyclic -- obvious variants are general or abelian difference sets.
$D=\{1,2,4\}\subset\Bbb{Z}_7$ is an example for a $(7,3,1)$ cyclic difference set.

Difference sets have been studied by mathematicians for more than half a century, see
\cite{baumert:71,jungnickel:92,pott:95} for a survey. They prove useful in various fields
like finite geometry or design theory. According to eq.~\myref{eq:C_N_k_lambda} it is
obvious that cyclic difference sets and binary sequences with {\em constant off-peak
autocorrelations} are essentially the same objects.

The fundamental questions in the field of difference sets concern their {\em existence} and
their {\em construction}. Albeit being unsolved in general, the existence question can 
be answered for special sets of parameters $(N,k,\lambda)$. Most of the results are
non-existence Theorems which impose necessary conditions on $(N,k,\lambda)$ for a cyclic 
difference set to  exist \cite{baumert:71}.
A simple non-existence Theorem is given by eq.~\myref{eq:constraint}, which
for a (general) difference set reads
\begin{equation}
  \label{eq:constraint_cds}
  k(k-1) = \lambda (N-1).
\end{equation}
See \cite{baumert:71} for other, more sophisticated non-existence theorems.
All known existence proofs are constructive, i.e.\ they provide an explicit
method to construct a cyclic difference set.

%% file: perfect.tex
\section{Perfect sequences}
\label{sec:perfect_sequences}

We define a {\em perfect sequence} to be a binary sequence with one-valued off-peak
autocorrelations of minimum amount, i.e.\ with
\begin{equation}
  \label{eq:C_perfect}
  C_{g\neq 0} = \left\{
    \begin{array}{rl}
      0 & N\equiv 0\bmod 4\\
      1 & N\equiv 1\bmod 4\\
      2 & N\equiv 2\bmod 4\\
      -1 & N\equiv 3\bmod 4\\
    \end{array}
  \right..
\end{equation}
Eq.~\myref{eq:constraint_cds} excludes the existence
of perfect sequences with all $C_{k>0} = -2$ for $N=4t-2>2$.

Obviously any perfect sequence is a ground state of the Bernasconi model.
To find perfect sequences, we have to look for 
cyclic difference sets with parameters
\begin{equation}
  \label{eq:desired_n}
  k-\lambda = t \qquad N=4t, 4t\pm1, 4t-2.
\end{equation}

\subsection{Hadamard difference sets}

Let us consider first the case $N\equiv3\bmod4$.
According to eq.~\myref{eq:constraint_cds}, the cyclic difference set that corresponds 
to a perfect sequence must have parameters
of the form
\begin{equation}
  \label{eq:hadamard_params}
  (N,k,\lambda) = (4t-1, 2t-1, t-1)
\end{equation}
for some integer $t>0$. Such difference sets are called {\em Hadamard difference sets}.
\label{sec:hadamard_cds}
\begin{table}[htb]
\begin{tabular}{ll}
$N = 2^j-1$, $j\geq 2$ & $m$-sequence \\
$N = 4t+3\quad$ prime & Legendre sequence \\
$N = p(p+2)\quad p,p+2$ prime & twin-prime sequence
\end{tabular}
\caption[Tab1]{\label{tab:hadamard}Classes of $N$ which allow the construction
of a Hadamard difference set resp.\ a perfect binary sequence with all 
off-peak
correlations $C_g=-1$.
}
\end{table} 
All {\em known} Hadamard difference sets can be classified into three classes according to the
value of $N$ (table \ref{tab:hadamard}). 
In fact it has been shown that all Hadamard
difference sets with $N \leq 10000$ belong to one of the above parameter classes, with
seventeen possible exceptions (all of them $>1000$)
\cite{song:golomb:94}. This result has been accomplished by the extensive use of non-existence
Theorems. A physicist may conclude that there are no Hadamard difference sets for other
parameters. The three classes cover 195 out of 250 values 
$N=4t+3<1000$. Table \ref{tab:hadamard} lists a construction-rule for every parameter class.
Beside these rules, there are additional construction
methods which lead to non-equivalent Hadamard difference sets for some parameter values
like $N = 4t+3 = 4x^2+27$, $N$ prime, for integer $x$ \cite{baumert:71}.

We start with the description of $m$-sequences.
Let $p$ be a prime and $F=GF(p)$ be the finite field of order $p$. 
%
A 
{\em shift register sequence } produced by a 
{\em linear feedback shift register  LFSR} of
order $n$ over $F$ is a sequence $\mathbf{a}=(a_k)$ of elements
from $F$ that satisfies the linear recurrence relation
\begin{equation}
  \label{eq:recurrence}
  a_i = \sum_{j=1}^n c_ja_{i-j} \qquad i \geq n.
\end{equation}
The sequence is uniquely determined by the initial conditions $(a_0,\ldots,a_{n-1})$
and the feedback coefficients $(c_1,\ldots,c_n)$, $c_j\in F$.
The name ``shift register sequence'' stems from a hardware realization of 
eq.~\myref{eq:recurrence}.

There are $p^n$ distinct $n$-tuples over $F$. The $\mathbf{0}$ tupel $(0,\ldots,0)$
reproduces itself under the linear recurrence relation. Therefore a 
LFSR 
of order $n$ over
$GF(p)$ produces a sequence which is ultimately periodic with period $\leq p^n-1$. 
A sequence with least period
$p^n-1$ is called a maximum period sequence or, for short, an $m$-sequence.

Note that in an $m$-sequence each $n$-tupel except $\mathbf{0}$ occurs exactly once. Hence an
$m$-sequence is independent of the initial conditions and solely determined by the 
recursion coefficients $\mathbf{c}$. A well known Theorem \cite[ch.6]{jungnickel:93} states
that a LFSR with coefficients $(c_1,\ldots,c_n)$ produces an $m$-sequence if and only if
the associated feedback polynomial
\begin{equation}
  \label{eq:feedback_poly}
  f(x) = 1 - c_1x - c_2x^2 - \cdots -c_nx^n
\end{equation}
is a primitive polynomial.  
%
Primitive polynomials of all degrees
exist for finite fields of any order $p$, i.e.\ for every prime $p$ there
is an LFSR of length $n$ that produces a sequence of period $p^n-1$, the maximum
value.

Supplied with an $m$-sequence 
over $F=\mathrm{GF}(2)$, 
we 
can define 
a binary sequence via
\begin{equation}
  \label{eq:m_to_binary}
  s_i = (-1)^{a_i} = (-1)^{c_1a_{i-1}+\dots+c_na_{i-n}}.
\end{equation}
This sequence then satisfies the multiplicative recursion
\begin{equation}
  \label{eq:s_recursion}
  s_i = s_{i-1}^{c_1}\cdot s_{i-2}^{c_2}\cdot \cdots \cdot s_{i-n}^{c_n}.
\end{equation}
which produces an $m$-sequence,  
meaning that 
every $n$-tupel of $\pm1$-values 
except the all-1 tupel $(1,1,\dots,1)$ 
occurs exactly once 
as a subsequence $(s_t, s_{t+1}, \ldots , s_{t+n-1})$ in $(s_i)$.  
It is easy to see that the product sequence $d_i = s_is_{i+g}$ has the same recursion as the $s_i$
\begin{equation}
  \label{eq:d_recursion}
  d_i = d_{i-1}^{c_1}\cdot d_{i-2}^{c_2}\cdot \cdots \cdot d_{i-n}^{c_n}.
\end{equation}
Since $(s_i)$ is an $m$-sequence, $(d_i)$ equals $(s_i)$ (for $g\neq 0$) 
except for a possible shift 
in the index,
\begin{equation}
  \label{eq:shift_index}
  d_i = s_{i+s}
\end{equation}
for some $s$. Using this result, we find for the correlations
of the sequence $S=(s_0, \ldots , s_{N-1})$, $N=2^n-1$,   
\begin{equation}
  \label{eq:m-correlations}
  C_g = \sum_{i=0}^{2^n-2}s_{i+s} = \sum_{i=0}^{2^n-2}s_i = -1
\end{equation}
since exactly 
$2^{n-1}$ of the $s_i$ equal $-1$ and the remaining $2^{n-1}-1$ equal $+1$.

As we have seen, $m$-sequences over $GF(2)$ with period $2^n-1$ can be used to
construct perfect binary sequences with the same period. This construction requires
a primitive polynomial of order $n$ over $GF(2)$. Tables of such polynomials can be 
found in the literature 
\cite{watson:62,zivkovic:94a,zivkovic:94b}.

Now we discuss the construction of {\em Legendre sequences}.
Let $p$ be an odd prime. Exactly half of the elements 
$g\in\textrm{GF}(p)^* = GF(p)\setminus\{0\}$ 
are {\em squares},
i.e.\ they have
a representation $g \equiv x^2\bmod p$ with 
$x\in\textrm{GF}(p)^*$. 
The other half cannot
be written as a square. A convenient notation for this property is the {\em quadratic character}
\begin{equation}
  \label{eq:def_chi}
  \chi_p(g) = \left\{
  \begin{array}{rl}
    0 & g = 0 \\
    1 & g \textrm{ is a square in }\textrm{GF}(p)^*\\
   -1 & \textrm{otherwise}
  \end{array}
  \right..
\end{equation}
The quadratic character can easily be calculated using Euler's criterion 
\cite[chap.~15]{schroeder:84}:
\begin{equation}
  \label{eq:euler}
  \chi_p(g) = g^{\frac12(p-1)}\bmod p,
\end{equation}
and it obeys 
\begin{equation}
  \label{eq:chi_ortho}
  \sum_{x\in\textrm{GF}(p)} \chi_p(x)\chi_p(x+g) = -1
\end{equation}
for all $g \neq 0$. This {\em orthogonality} of $\chi_p$ translates into
good correlation properties of the so called {\em Legendre sequence} of length $p$
\begin{equation}
  \label{eq:legendre_seq}
  s_i = \left\{
    \begin{array}{cl}
          1     & i \equiv 0 \bmod p \\
      \chi_p(i) & \textrm{otherwise}
    \end{array}
  \right..
\end{equation}
The periodic off-peak autocorrelations of the Legendre sequence read
\begin{eqnarray*}
  C_g & = & \sum_{i=0}^{p-1} s_is_{i+g} \\
      & = & \chi_p(g) + \chi_p(-g) + \sum_{i=0}^{p-1} \chi_p(i)\chi_p(i+g) \\
      & = & \chi_p(g)\left(1+(-1)^{(p-1)/2}\right) - 1,
\end{eqnarray*}
i.e.
\begin{equation}
  \label{eq:legendre_C}
  C_g = \left\{
    \begin{array}{rl}
      -1 & p \equiv 3 \bmod 4 \\
      +1,-3 & p \equiv 1 \bmod 4
    \end{array}
  \right..
\end{equation}
For $p \equiv 3 \bmod 4$ the Legendre sequence corresponds to a 
cyclic Hadamard difference set. For
$p \equiv 1 \bmod 4$ exactly half of the autocorrelations take on the value $+1$, the other half $-3$.

The discrete Fourier transform (DFT),
\begin{equation}
  \label{eq:fourier_def}
  B_k = \frac{1}{\sqrt{N}}\sum_{j=0}^{N-1} e^{-i2\pi kj/N}s_j.
\end{equation}
of a Legendre sequence is given by \cite[chap.~15]{schroeder:84}
\begin{equation}
  \label{eq:fourier_legendre}
  B_k = \left\{
    \begin{array}{ll}
      \frac{1}{\sqrt{p}} - is_k & m \not\equiv 0 \bmod p \\
      \frac{1}{\sqrt{p}} & m \equiv 0 \bmod p
    \end{array}
  \right.
\end{equation}
for $p\equiv 3 \bmod 4$.
This peculiar similarity between the Legendre sequence and its
DFT, $\mathfrak{Im}B_k = -s_k$ $(k\not\equiv 0\bmod p)$, turns the
Legendre sequences into ground states of another deterministic
model with glassy properties,
\begin{equation}
  \label{eq:def_sine_model}
  \begin{split}
    \tilde H &= \sum_{k=1}^{p-1}|s_k + \mathfrak{Im}B_k|^2 \\
             &= \sum_{k=1}^{p-1}|s_k - \frac{1}{\sqrt{p}}\sum_{j=0}^{p-1}\sin(
                2\pi jk/p) s_j|^2,
  \end{split}
\end{equation}
the so called sine-model \cite{marinari:parisi:ritort:94b}.

The last entry in table \ref{tab:hadamard} refers to the {\em twin-prime sequences}.
Let $p,q$ be odd primes with $p > q$. Consider the binary sequence of length $pq$,
\begin{equation}
  \label{eq:two_prime_seq}
  s_i = \left\{
    \begin{array}{cl}
      +1 & i \equiv 0 \bmod p\\
      -1 & i \equiv 0 \bmod q, 
                    i\not\equiv 0 \bmod p\\
      \chi_p(i)\chi_q(i) & \textrm{otherwise}
    \end{array}
  \right..
\end{equation}
If $p$ and $q$ are twin primes ($p = q+2$), all correlations of this sequence equal $-1$.
A proof will be given in section \ref{sec:beyond_perfect_sequences}, 
where we discuss the case of general
odd prime numbers $p$ and $q$.

\subsection{Menon difference sets}
\label{sec:menon_cds}

Now we consider the case $N\equiv0\bmod4$. 
Eq.~\myref{eq:constraint_cds} enforces 
\begin{equation}
  \label{eq:menon}
  (N,k,\lambda) = (4u^2, 2u^2-u, u^2-u)
\end{equation}
for some integer $u>0$ for the cyclic difference set that corresponds to a perfect
sequence. A cyclic difference set that satisfies eq.~\myref{eq:menon} is called
{\em Menon difference set}. The corresponding binary sequence 
forms the row of a {circulant Hadamard\footnote{This is the reason, why Menon difference sets
are sometimes called Hadamard difference sets in the mathematical literature. Our Hadamard
difference sets are then called Paley-Hadamard difference sets 
\cite[p.\ 232]{jungnickel:93}.} matrix.
Example for $u = 1$, written as binary sequence:
\begin{equation}
  \label{eq:menon4}
  (s) = (+1,+1,+1,-1).
\end{equation}
Unfortunately, this is the only known example of a Menon difference set.
In fact, it has been shown \cite{schmidt:97} that no
Menon difference set exists
for $1 < u < 250$ with the possible exceptions
$u=165$ and $u=231$. There is a tight connection between Menon difference sets
and {\em Barker sequences}, i.e.\ binary sequences with
\begin{equation}
  \label{eq:barker}
  \left|\sum_{i=0}^{N-1-g}s_is_{i+g}\right| \leq 1
\end{equation}
The existence of a Barker sequence of even length $N$ implies the existence of
a Menon difference set of order $N$ \cite{turyn:storer:61,turyn:68}. It is not known whether
the converse is also true. It has been shown, however, that there are no Barker sequences
for $13 < N < 2.5\cdot10^9$ with the possible exception $N=1596961444$ \cite{schmidt:97}.
To summarize: It is a still unproven, but widely accepted conjecture, 
that no Menon difference set exists for $N>4$.

\subsection{Other difference sets}
\label{sec:singer_cds}

Another family of cyclic difference sets can be constructed for parameters
\begin{equation}
  \label{eq:singer}
  (N,k,\lambda) = \left(\frac{q^{\nu+1}-1}{q-1}, \frac{q^\nu-1}{q-1}, 
                        \frac{q^{\nu-1}-1}{q-1}\right)
\end{equation}
where $\nu \leq 1$ and $q$ is a prime power.
Such sets are called {\em Singer difference sets} \cite{baumert:71}.

For $q=2$ we rediscover the parameters of the Hadamard difference sets based on 
$m$-sequences, $q=3, \nu=1$ corresponds to the lonely Menon difference
set \myref{eq:menon4}. The only other cases where Singer difference sets yield
ground states of the Bernasconi model are
\begin{center}
\begin{tabular}{rrrr}
 $q$ & $\nu$ & $(N,k,\lambda)$ & $C_{g\neq0}$ \\\hline
 4 & 1 & $(5,1,0)$ & 1 \\
 5 & 1 & $(6,1,0)$ & 2 \\
 3 & 2 & $(13,4,1)$ & 1
\end{tabular}
\end{center}

\noindent
Since the Singer difference sets yield ground states for only 3 new values
of $N$, we will not discuss their construction.

According to eq.~\myref{eq:constraint_cds}, a difference set with 
$N\equiv 1 \bmod 4$ and $C_g = 1$ must have parameters of the
form
\begin{equation}
  \label{eq:1mod4}
  (N,k,\lambda) = (2u(u+1)+1, u^2, \frac12u(u-1))
\end{equation}
for some integer $u > 0$. 
It has been shown \cite{eliahou:kervaire:92} that no such cyclic difference set exists 
for $3\leq u \leq 100$, i.e.\ $13 < N \leq 20201$, so again 
it is well-founded to believe 
that beyond $N=13$ there are no such sets at all.

For $N=4t-2$, sequences with
all $C_{k>0} = 2$ require 
\begin{equation}
  \label{eq:constraint_2mod4}
  3t = u^2+2
\end{equation}
with integer $u$, i.e.\ $N=2,6,22,26,66,\ldots$. Except for $N=2$ and $N=6$, no such sequences
are known.

Other difference sets do exist, but either their parameters coincide with
one of the sets already discussed, or they lead to values for the off-peak
autocorrelations that are far from a ground state.

%% file: almost.tex
\section{Beyond perfect sequences}
\label{sec:beyond_perfect_sequences}

In the preceeding section we have seen that for many values of $N$ perfect sequences
do not exist. In this section we introduce a generalization called {\em almost perfect sequences}
and discuss their appearance as groundstates of the Bernasconi model. In addition we present
a construction method for $N=pq$, $p$ and $q$ odd primes, that yields
configurations with low energies, and we discuss generalized {\em Jacobi sequences}.

\subsection{Almost perfect sequences}

An almost perfect sequence is a binary sequence with two-valued off-peak
autocorrelations of minimum amount, i.e.\ with
\begin{equation}
  \label{eq:C_almost_perfect}
  C_{g\neq 0} \in \left\{
    \begin{array}{rl}
      \{0,\pm4\} & N\equiv 0\bmod 4\\
      \{1,-3\} & N\equiv 1\bmod 4\\
      \{-2,2\} & N\equiv 2\bmod 4\\
      \{-1,3\} & N\equiv 3\bmod 4\\
    \end{array}
  \right..
\end{equation}
Again we have an ambiguity: for $N\equiv 0\bmod 4$ all correlations are either $\in\{0,+4\}$ or
$\in\{0,-4\}$. A sequence with three-valued correlations $C_g\in\{-4,0,+4\}$ does not match our
definition.

An almost perfect sequence corresponds to a set $D$ with two-valued replication number
where the two replication numbers differ by one:
\begin{equation}
  \label{eq:almost_lambda}
  \lambda(g) = \left\{
    \begin{array}{cl}
    k & g = 0 \\
    \lambda+1 & g \in U \subset G \\
    \lambda & g \notin U
    \end{array}
    \right..
\end{equation}
$U$ can be any subset of the underlying cyclic group $G$. 
We call a set $D$ with such replication number an
{\em almost cyclic difference set}.

Values of $N$ that allow the construction of perfect sequences are rare, as we have seen in the 
preceeding section. The question is 
whether 
the weaker constraint of almost perfectness can be 
fulfilled for more values of $N$.

\begin{figure}[htb]
  \includegraphics[width=\columnwidth]{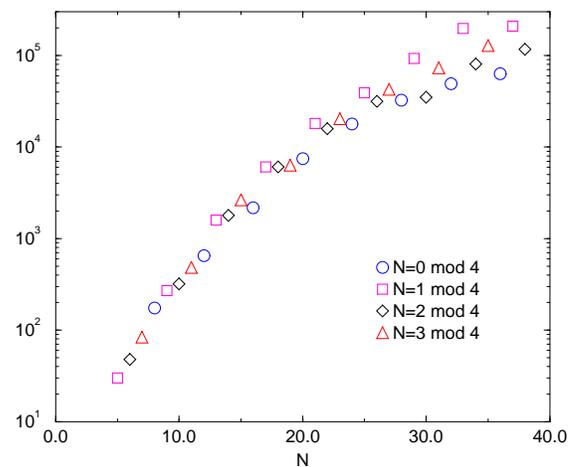}
        \caption[Fig1]{\label{fig:n_almost}Total number of almost perfect binary sequences.}
\end{figure}

To study this question we counted all almost perfect sequences of given length $N$ by exhaustive
enumeration (Figure \ref{fig:n_almost}) Their number 
seems to increase exponentially with $N$ but much slower than $2^N$. From figure \ref{fig:n_almost}
we may conclude that
that there are almost perfect sequences for {\em all} values of $N$ but that their
relative volume in configuration space decreases exponentially with $N$.

For larger values of $N$, this conclusion is supported by a heuristic numerical search.
We start with a random sequence and lower its energy by single spin-flips
until we get a sequence whose energy can no longer be lowered by flipping a single spin.
If this local minimum is an almost perfect sequence, the procedure stops. If not,
it starts again with a new random
configuration. 

With this algorithm we found almost perfect sequences
for all $N<84$. The CPU time increased strongly with $N$,
so we stopped the search for larger values of $N$ before an
almost perfect sequence had been found. Random search is of course
not well suited for configurations that are very rare. As an example, 
consider this: Nine workstations searched the configuration space for $N=86$
in parallel to find a sequence with $C_g=\pm2$. After 20 days, two machines
have found sequences with all $C_g=\pm2$ for all $g>0$
except $C_{43}=6$, while
the other 7 machines still report sequences with at least two correlations
$C_g=6$ as their best results.
The failure of random search for large values of $N$ does not indicate the 
non-existence of almost perfect sequences.

Compared to difference sets, much less is known about the construction of almost difference
sets. We only know two construction methods for almost perfect sequences:
Legendre sequences for $N=4t+1$ prime (see sec.~\ref{sec:perfect_sequences})
and a construction due to Lempel
et al.\ \cite{lempel:cohn:eastman:77} 
for $N=p^m-1$ where $p$ is an odd prime and $m$ is a positive integer. 

The construction of Lempel et al.\ works like this: Consider the finite field
$F=GF(p^m)$. Let $G$ denote the multiplicative group of $F$ and let $\alpha$ be any 
primitive element
of $F$. The subset $D$ of $G$ defined by
\begin{equation}
  \label{eq:def_D_Lempel}
  D = \{\alpha^{2i+1}-1\}_{i=0}^{k-1}
\end{equation}
where
\begin{equation}
  \label{eq:k_Lempel}
  k = \frac12(p^m-1)
\end{equation}
is an almost difference set (see \cite{lempel:cohn:eastman:77} for a proof). 
The binary sequence that corresponds to 
$D$ according to equation \myref{eq:SfromD_mult} is an almost perfect
sequence with correlations
\begin{equation}
  \label{eq:C_Lempel}
  C_g = \left\{
    \begin{array}{rl}
      \textrm{$2$ or $-2$} & \textrm{if $k$ is odd} \\
      \textrm{$0$ or $-4$} & \textrm{if $k$ is even} \\
    \end{array}
  \right..
\end{equation}
Example for $p=7$ and $m=1$: $GF(7)$ is equivalent to $\Bbb{Z}_7$, the integers $0,\ldots,6$
with addition and multiplication modulo $7$. The primitive element $\alpha=3$ leads to
\begin{equation}
  \label{eq:Lempel_ex_D}
  D = \{2,5,4\}
\end{equation}
and the corresponding binary sequence reads
\begin{equation}
  \label{eq:Lempel_ex_S}
  S = (-1,-1,+1,-1,+1,+1)
\end{equation}
with correlations $C_1=C_2=C_4=C_5=-2$ and $C_3=+2$.

For $N\equiv2\bmod4$, any almost perfect sequence is a ground state of the Bernasconi
model, i.e.\ the  construction of Lempel et al.\ yields ground states for
$N=4t+2=p^m-1$, where $p$ is an odd prime. Since our numerical results suggests that almost perfect
sequences do exist for all values of $N$, we arrive at the conjecture, that the
ground states of the Bernasconi model for $N=4t+2$ are given by almost perfect sequences
with ground state energy $E=4(N-1)$. This conjecture is in agreement with former numerical
results \cite{marinari:parisi:ritort:94a}.  

For the three other residue classes, the question whether an almost perfect sequence is a ground 
state depends on $|U|$, i.e.\ on the number of correlations that deviate from the optimum value.

Consider first the case $N=4t+1$ where $t$ is a positive integer. 
Let the value $(-1)^i(2i+1)$ appear $2n_i$ times among the
correlations of a sequence, $i = 0,1,\ldots$. The factor $2$ reflects the
fact, that $C_g=C_{N-g}$ must hold. The possible excitations above the
theoretical minimum energy $E=4t$ are given by
\begin{eqnarray}
  \label{eq:Delta_E_1}
  \Delta E & = & \frac1{16}(E-4t) = \frac12\sum_{i=1}^\infty i(i+1)n_i \\
  & = & n_1 + 3n_2 + 6n_3 + 10n_4 + 15n_5 + \cdots. \nonumber
\end{eqnarray}
Since 
$\sum_{i=0}^{\infty}n_i = 2t$ 
must hold, the above sum is finite, of course.
Equation \myref{eq:constraint} imposes the constraint 
\begin{equation}
  \label{eq:constraint_1}
  \sum_{j=1}^\infty j(n_{2j}-n_{2j-1}) = \frac12u(u+1)-t
\end{equation}
on the $n_i$, where $u$ is a non-negative integer. An almost perfect
sequence has $n_i=0$ for $i > 1$ and equation \myref{eq:constraint_1} reduces
to
\begin{equation}
  \label{eq:constraint_almost_1}
  n_1 = t-\frac12u(u+1).
\end{equation}
This limits the possible values for $n_1$ to
\begin{equation}
  \label{eq:possible_n_1}
  n_1 \in \{t, t\!-\!1, t\!-\!3,\ldots,t\!-\!\frac12u_{\mathrm{max}}(u_{\mathrm{max}}+1)\}
\end{equation}
with
\begin{equation}
  \label{eq:max_u_1}
  u_{\mathrm{max}}(t) = \left\lfloor \frac12\sqrt{1+8t}\right\rfloor.
\end{equation}

\begin{table}[htb]
\begin{tabular}{r|l|r|c}
$N$ & $n_1$: possible values & found & ground state\\\hline
 5 & 0,1 &  0 & yes\\
 9 & 1,2 &  1 & yes\\
13 & 0,2,3 &  0 & yes\\
17 & 1,3,4 &  3 & yes\\
21 & 2,4,5 &  2 & yes\\
25 & 0,3,5,6 &  3 & yes\\
29 & 1,4,6,7 &  4 & yes\\
33 & 2,5,7,8 &  2 & yes\\
37 & 3,6,8,9 &  3 & yes\\
41 & 0,4,7,9,10 &  4 & yes\\
45 & 1,5,8,10,11 &  5 & a.\ p. \\
49 & 2,6,9,11,12 &  6 & a.\ p. \\
53 & 3,7,10,12,13 &  7 & ? \\
57 & 4,8,11,13,14 &  8 & ? \\
61 & 0,5,9,12,14,15 &  9 & ? \\
65 & 1,6,10,13,15,16 & 10 & ? \\
69 & 2,7,11,14,16,17 & 11 & ? \\
73 & 3,8,12,15,17,18 & 12 & ? \\
77 & 4,9,13,16,18,19 & 18 & ? \\
81 & 5,10,14,17,19,20 & 19 & ?
\end{tabular}
\caption[Tab1]{
  \label{tab:almost_1}
  Number of correlations $C_g = -3$ in low energy, almost perfect sequences for $N=4t+1$ 
  found by numerical search. An entry a.p.\ in column ``ground state'' indicates that any
  hypothetical sequence with lower energy must be an almost perfect sequence. The energy of the
  sequences is given by $E=16n_1+N-1$.
}
\end{table}

We ran a numerical search for low energy, almost perfect sequences up to size $N=81$.
Table \ref{tab:almost_1} shows the result. For $N\leq 41$ we perform an exhaustive enumeration
to check that the sequences we found are true ground states. Using Eq.~\myref{eq:constraint_1},
we can assert that any hypothetical sequence with energy below the value 
listed 
in 
table~\ref{tab:almost_1} must be an almost perfect sequence for $N=45$ and $N=49$.
This is not possible for larger values like $N=53$, 
where a configuration with $n_1=0$ and $n_2=1$ has lower energy ($\Delta E=3$) and satisfies
Eq.~\myref{eq:constraint_1}.
For $N=5$ and $N=13$, perfect sequences based on Singer difference sets are ground states.
Note that the Legendre sequences for $N\equiv1\bmod4$ prime
have $n_1=\frac14(N-1)$. This value is too large for a ground state, at least for those values
of $N$ considered in Table \ref{tab:almost_1}. 

Based on their numerical 
results for $N\leq 50$, Marinari et al. \cite{marinari:parisi:ritort:94a}
observed that
\begin{equation}
  \label{eq:marinari_1}
  \Delta E = t - 6 
\end{equation}
holds for $t=8,9,\ldots,12$. As can be seen from Table \ref{tab:almost_1}, this equation
seems to hold at least up to $t=18$ ($N=73$).

Consider now the case $N=4t+3$. This time, $2n_i$ denotes the number of correlations with
value $(-1)^{i+1}(2i+1)$. The excitation energy reads
\begin{eqnarray}
  \label{eq:Delta_E_3}
  \Delta E & = & \frac1{16}(E-4t-2) = \frac12\sum_{i=1}^\infty i(i+1)n_i \\
  & = & n_1 + 3n_2 + 6n_3 + 10n_4 + 15n_5 + \cdots. \nonumber
\end{eqnarray}
and Eq.~\myref{eq:constraint} leads to the constraint 
\begin{equation}
  \label{eq:constraint_3}
  \sum_{j=1}^\infty j(n_{2j}-n_{2j-1}) = \frac12u(u+1),
\end{equation}
where $u$ is a non-negative integer. 
An almost perfect sequence must have
\begin{equation}
  \label{eq:possible_n_3}
  n_1 \in \{0, 1, 3, 6, \ldots, \frac12u_{\mathrm{max}}(u_{\mathrm{max}}+1)\}
\end{equation}
with
\begin{equation}
  \label{eq:max_u_3}
  u_{\mathrm{max}}(t) = \left\lfloor \frac12\left(\sqrt{9+16t}-1\right)\right\rfloor.
\end{equation}

\begin{table}[htb]
\begin{tabular}{r|l|r|c}
$N$ & $n_1$: possible values & found & ground state\\\hline
27 & 1,3,6,10 & 3 & yes\\
39 & 1,3,6,10,15 &  3 & yes\\
51 & 1,3,6,10,15,21 &  6 & ?\\
55 & 1,3,6,10,15,21 &  10 & ?
\end{tabular}
\caption[Tab1]{
  \label{tab:almost_3}
  Number of correlations $C_g = 3$ in low energy, almost perfect sequences for $N=4t+3$ 
  found by numerical search. For all other values of $N=4t+3<75$, ground states are given
  by perfect sequences based on Hadamard difference sets.
}
\end{table}
 
Most values $N = 4t+3 < 75$ allow the construction of Hadamard cyclic difference sets,
i.e.\ perfect sequences. Table \ref{tab:almost_3} lists low energy sequences found by 
numerical search for the 4 remaining values $N < 75$. 
The numerical search for $N=75$
yielded 
low 
energy configurations $\Delta E=14$ with $(n_1,n_2)=(8,2)$ and $(n_1,n_2)=(11,1)$.
This does of course not exclude the possibility of an almost perfect sequence with lower energy.

\subsection{The case $N\equiv 0\bmod 4$}

So far all known ground states for $N$ odd 
or 
$N\equiv2\bmod4$ are perfect or almost 
perfect sequences. The first exceptions occur for $N=4t$.
Let $m_i$ ($n_i$) denote the number of correlations with value $4i$ ($-4i$).
The excitation above the theoretical minimum $E=0$ reads
\begin{equation}
  \label{eq:Delta_E_0}
  \Delta E = \frac1{16}E = \sum_{i=1}i^2(m_i+n_i)
\end{equation}
and Eq.~\myref{eq:constraint} leads to the constraint 
\begin{equation}
  \label{eq:constraint_0}
  \sum_{i=1}^\infty i(m_i-n_i) = u^2-t,
\end{equation}
where $u$ is a non-negative integer. 

\begin{table}[htb]
\begin{center}
\begin{tabular}{r|r|rrrr|c}
$N$ & $\Delta E$ & $m_2$ & $m_1$ & $n_1$ & $n_2$ & ground state\\\hline
 8 & 1 & 0 & 0 & 1 & 0 &  yes \\
12 & 1 & 0 & 1 & 0 & 0 &  yes \\
16 & 3 & 0 & 0 & 3 & 0 &  yes \\
20 & 4 & 0 & 0 & 4 & 0 &  yes \\
   &   & 0 & 4 & 0 & 0 &  yes \\
24 & 2 & 0 & 0 & 2 & 0 &  yes \\
28 & 5 & 0 & 1 & 4 & 0 &  yes \\
32 & 5 & 0 & 3 & 2 & 0 &  yes \\
36 & 4 & 0 & 2 & 2 & 0 &  yes \\
40 & 5 & 0 & 2 & 3 & 0 &  yes \\
44 & 9 & 0 & 7 & 2 & 0 &  ? \\
48 & 6 & 1 & 2 & 0 & 0 &  ? \\
52 & 8 & 0 & 2 & 6 & 0 &  ? \\
56 & 12 & 1 & 4 & 4 & 0 &  ? \\
60 & 7 & 0 & 4 & 3 & 0 &  ? \\
64 & 14 & 0 & 6 & 4 & 1 &  ? \\
68 & 16 & 0 & 12 & 4 & 0 &  ? \\
72 & 13 & 0 & 10 & 3 & 0 &  ? \\
76 & 17 & 0 &  7 & 10 & 0 & ?
\end{tabular}
\end{center}
\caption[Tab1]{
  \label{tab:almost_0}
  Low energy configurations for $N=4t$, found by exhaustive enumeration ($N \leq 40$)
  or numerical search.
}
\end{table}

Table \ref{tab:almost_0} shows low energy configurations in terms of $m_i$ and $n_i$, found
either by exhaustive enumeration ($N\leq 40$) or numerical search. For $N=28, 32, 36$ and $40$,
the ground states are not given by almost perfect sequences, but by sequences which still 
obey
$|C_g| \leq 4$. For $N=48, 56, 64$, even these bounds seem to be violated.

The construction of Lempel et al.\ \cite{lempel:cohn:eastman:77} for $N=p^m-1$, $p$ odd prime,
yields almost perfect sequences with $n_1 = N/4$. These sequences are not ground states for all
values of $N$ in Table \ref{tab:almost_0}. They do not even have the lowest energy among all
almost perfect sequences.

Wolfmann \cite{wolfmann:92,pott:bradley:95} describes a construction for $N\equiv0\bmod4$, that yields
sequences with all $C_g=0$, except $C_{N/2}=-(N-4)$. The energy of such sequences is 
much higher than the ground state energy, at least for those $N$ considered in 
Table \ref{tab:almost_0} (except $N=8$).

\subsection{Two-prime sequences}
\label{sec:two-prime_series}

In this section, we describe the construction of low autocorrelation binary sequences for
$N=pq$, $p$ and $q$ prime, which can be shown to be true ground states of the Bernasconi model
if $p$ and $q$ are twin primes. For general primes $p$ and $q$, they have at least a low
energy, far below the energies that can be found by numerical search for large $N$.

Let $p$ and $q$ be odd primes and $p > q$. Then
$G = \Bbb{Z}_q \times \Bbb{Z}_p$ is a $pq$-element cyclic (additive) 
group. Consider the subset $D \subset G$ given by
\begin{equation}
  \begin{split}
  \label{eq:Dpq_definition}
  D = &\underbrace{\{(a, b)\in G \,|\, a, b\neq 0, \chi_q(a) = \chi_p(b)\}}_{=: M}\\ 
  & \cup \{(a, 0)\in G \,|\, a \in \Bbb{Z}_q \}
  \end{split}
\end{equation}
$D$ has 
\begin{equation}
  \label{eq:Dpq_k}
  k = \frac12 (p-1)(q-1) + q
\end{equation}
elements and $M$ is a multiplicative group. 

To calculate the correlations of the two-prime sequence
\begin{equation}
  \label{eq:2-prime-series}
  s_i = \left\{
  \begin{array}{rl}
    +1 & (i \bmod q, i \bmod p) \in D \\
    -1 & \mathrm{otherwise}
  \end{array}
  \right.,
\end{equation}
we generalize the proof given in \cite{beth:etal:85} to arbitrary primes
$p$ and $q$. Note that the sequences given by Eqs.~\myref{eq:two_prime_seq} and 
\myref{eq:2-prime-series} are the same.

Let $\lambda(x,y)$ denote the
number of different representations
\begin{equation}
  \label{eq:Dpq_representation}
  (x,y) = (a_1,b_1) - (a_2,b_2) \qquad (a_i,b_i)\in D
\end{equation}
of $(x,y)\in G-\{0\}$.
Since $MD = D$, differences $(x,y)$ and
$(m_1x,m_2y)$ with $(m_1,m_2)\in M$ will occur equally often in $D$. Hence there are
constants $\lambda_1$, $\lambda_2$, $\lambda_3$ and $\lambda_4$ such that
\begin{equation}
  \label{eq:Dpq_lambda_x_y}
  \lambda(x,y) = \left\{
    \begin{array}{ll}
      \lambda_1 & \textrm{ for } x,y \neq 0 \textrm{ and } \chi_q(x) = \chi_p(y)\\
      \lambda_2 & \textrm{ for } x,y \neq 0 \textrm{ and } \chi_q(x) \neq \chi_p(y)\\
      \lambda_3 & \textrm{ for } x = 0\\
      \lambda_4 & \textrm{ for } y = 0\\
    \end{array}
  \right.  
\end{equation}
The corresponding autocorrelations read
\begin{equation}
  \label{eq:Dpq_C}
  C(x,y) = 4\lambda(x,y) -pq + 2(p-q) - 2.
\end{equation}

To calculate $\lambda_{1}, \ldots\lambda_4$, 
we will look closely at the number of
difference representations in $\Bbb{Z}_t^*$, the group
of integers $1,\ldots,t-1$ with multiplication mod $t$, $t$ prime.
Let $\varepsilon_1, \varepsilon_2\in\{1,-1\}$ and $x\in\Bbb{Z}_t^*$ and
\begin{multline}
  \label{eq:lambda_eps_def}
  \lambda_{\varepsilon_1,\varepsilon_2}^t(x) = |\{(a_1, a_2) | x = a_1 - a_2, \\
  a_1, a_2 \in\Bbb{Z}_t^*, \, \chi_t(a_i) = \varepsilon_i\}|.
\end{multline}
Then we have
\begin{equation}
  \label{eq:lambda_eps_1}
  \lambda_{\varepsilon_1,\varepsilon_2}^t (x) = \left\{
  \begin{array}{ll}
    \lambda_{\varepsilon_1,\varepsilon_2}^t (1) & \text{if } \chi_t(x) = 1 \\
    \lambda_{-\varepsilon_1,-\varepsilon_2}^t (1) & \text{if } \chi_t(x) = -1
  \end{array}
  \right..
\end{equation}
We define $\lambda_{\varepsilon_1,\varepsilon_2}^t = \lambda_{\varepsilon_1,\varepsilon_2}^t(1)$.
We also write $+,-$ instead of $1,-1$. Then clearly
\begin{equation}
  \label{eq:lambda_def_ab}
  \lambda_{++}^t + \lambda_{+-}^t + 1 =  \lambda_{--}^t + \lambda_{-+}^t = \frac{t-1}2.
\end{equation}
From eq.~\myref{eq:chi_ortho} we deduce 
\begin{equation}
  \label{eq:lambda_def_c}
  \lambda_{++}^t + \lambda_{--}^t- \lambda_{+-}^t - \lambda_{-+}^t 
  = -1.
\end{equation}
Furthermore, for $\varepsilon\in\{1, -1\}$ we set
\begin{equation}
  \label{eq:def_delta_t}
  \delta_\varepsilon^t = \left\{
  \begin{array}{ll}
    1 & \text{if $\chi_t(-1) = \varepsilon$} \\
    0 & \text{else}
  \end{array}
  \right..
\end{equation}
We decompose $\Bbb{Z}_t^*$ in intervals of squares resp.\ nonsquares only, i.e.\
\begin{equation}
  \label{eq:decompose_Z}
  \Bbb{Z}_t^* = \bigcup_{i=1}^r [m_{i-1}+1,\ldots,m_i], \quad m_0 = 0,\, m_r = t-1
\end{equation}
where $[m_{i-1}+1,\ldots,m_i]$ consists of squares for odd $i$ and of nonsquares for
even $i$.
Then $\lambda_{-+}^t$ resp.\ $\lambda_{+-}^t$ counts the difference representations
$1=(m_i+1)-m_i$ for $i$ odd resp.\ $i$ even, and thus
\begin{equation}
  \label{eq:lambda_def_d}
  \lambda_{-+}^t - \lambda_{+-}^t =  \delta_{-}^t.
\end{equation}
The above equations for the parameters $\lambda_{\varepsilon_1,\varepsilon_2}^t$ provide an
easily solvable system of linear equations with solution
\begin{equation}
  \label{eq:lambda_eps_solution}
  \begin{split}
  \lambda_{++}^t &= \frac{t-1}4 + \frac{\delta_{-}^t}2 - 1 = \left[\frac{t-3}4\right] \\
  \lambda_{--}^t = \lambda_{+-}^t &= \frac{t-1}4 - \frac{\delta_{-}^t}2 = 
                                   \left\lceil \frac{t-3}4\right\rceil \\
  \lambda_{-+}^t &= \frac{t-1}4 + \frac{\delta_{-}^t}2 = \left[\frac{t+1}4\right]
  \end{split}.
\end{equation}
Finally, we define
\begin{equation}
  \label{eq:def_delta_pq}
  \begin{split}
  \delta^{p,q} &=
  \left\{ \begin{array}{ll}
    1 & \text{if $\chi_p(-1) = \chi_q(-1)$} \\
    0 & \text{else}
  \end{array}\right.\\
  &=
  \left\{ \begin{array}{ll}
    1 & \text{if $p\equiv q\bmod 4$} \\
    0 & \text{else}
  \end{array} \right..
  \end{split}
\end{equation}
For the value $\lambda_1$ we have:
\begin{eqnarray}
  \label{eq:lambda_1_long}
  \lambda_1 &=& \lambda(1,1) \\\nonumber
            &=& \lambda_{+-}^q(\lambda_{+-}^p+1+\delta_{-}^p) +\lambda_{-+}^q(\lambda_{-+}^p+\delta_{+}^p) +\\
            & & \lambda_{++}^q(\lambda_{++}^p+1+\delta_{+}^p) +\lambda_{--}^q(\lambda_{--}^p+\delta_{-}^p) + 1
                + \delta^{p,q} \nonumber
\end{eqnarray}
where in the final term $1 + \delta^{p,q}$ the contribution 1 counts the difference representation 
$(1,1)=(1,1)-(0,0)$ and $\delta^{p,q}$ counts $(1,1)=(0,0)-(-1,-1)$. Substituting the parameter values
obtained above, we then obtain
\begin{equation}
  \label{eq:lambda_1_result}
  \lambda_1 = \frac14(pq-2(p-q)+1) + \frac{\delta^{p,q}}2.
\end{equation}
Let $z\in\Bbb{Z}_p^*$ with $\chi_p(z)=-1$. We then obtain the value $\lambda_2$ as:
\begin{eqnarray}
  \label{eq:lambda_2_long}
  \lambda_2 &=& \lambda(1,z)\nonumber \\
            &=& \lambda_{-+}^q(\lambda_{+-}^p+1+\delta_{-}^p) +\lambda_{+-}^q(\lambda_{-+}^p+\delta_{+}^p) +\\
            & & \lambda_{--}^q(\lambda_{++}^p+1+\delta_{+}^p) +\lambda_{++}^q(\lambda_{--}^p+\delta_{-}^p) + 1
                - \delta^{p,q} \nonumber
\end{eqnarray}
where the final term $1- \delta^{p,q}$ counts the difference representation $(1,z) = (0,0)-(-1,-z)$.
Substituting our parameter values gives
\begin{equation}
  \label{eq:lambda_2_result}
  \lambda_2 = \frac14(pq-2(p-q)+1) - \frac{\delta^{p,q}}2.
\end{equation}
Furthermore,
\begin{eqnarray}
  \label{eq:lambda_3_long}
  \lambda_3 &=& \lambda(0,1) \nonumber\\
            &=& (\lambda_{++}^p +\lambda_{--}^p)\frac{q-1}2 + q-1\nonumber\\
            &=& \frac14(p+1)(q-1)
\end{eqnarray}
and
\begin{eqnarray}
  \label{eq:lambda_4_long}
  \lambda_4 &=& \lambda(1,0) \nonumber\\
            &=& (\lambda_{++}^q +\lambda_{--}^q)\frac{p+1}2 + \lambda_{+-}^q + \lambda_{-+}^q + 2\nonumber\\
            &=& \frac14(q-3)(p-1) + q.
\end{eqnarray}
The correlations for $g \neq 0$ finally read
\begin{equation}
  \label{eq:correlations_two_prime}
  C_g = \left\{
  \begin{array}{ll}
    -1 + 2\delta^{p,q} & \text{for $\chi_q(g) = \chi_p(g)$, 
                              $g \not \equiv 0 \bmod p,q$} \\
    -1 - 2\delta^{p,q} & \text{for $\chi_q(g) \neq \chi_p(g)$
                                $g \not \equiv 0 \bmod p,q$}\\
    p - q - 3 & \text{for $g \equiv 0 \bmod q$} \\
    q - p + 1 & \text{for $g \equiv 0 \bmod p$} 
  \end{array}
  \right.
\end{equation}
with
\begin{equation}
  \label{eq:energy_two_prime}
  \begin{split}
  E = &(1+4\delta^{p,q})(p-1)(q-1) +\\
      &(p-q-3)^2(p-1) + (p-q-1)^2(q-1).
  \end{split}
\end{equation}
For $p-q=2$, all correlations are $-1$, i.e.\ we proved that
the twin-prime sequence Eq.~\myref{eq:two_prime_seq} corresponds
to a cyclic Hadamard difference set. For $p-q>2$, the construction does
not necessarily lead to a ground state, as can be seen for $N=21, 33, 57, 65, 69$ 
(table \ref{tab:almost_1}) and $N=39, 51, 55$ (table \ref{tab:almost_3}).
Nevertheless two-prime sequences have energies well below those
found by extensive numerical search for large $N=pq$ and small difference $p-q$.

\subsection{Jacobi sequences}

The {\em Jacobi
symbol} $\psi_N(j)$ is a generalization of the quadratic character $\chi_p(j)$ 
(Eq.\ \ref{eq:def_chi}) to the case that $N=p_1^{r_1}p_2^{r_2}\cdots p_s^{r_s}$ is the product of 
odd primes:
\begin{equation}
  \label{eq:def_jacobi}
  \psi_N(j) = \prod_{i=1}^s \chi_{p_i}^{r_i}(j).
\end{equation}
Note that $\psi_N(j) = 0$ if $(j,p_i)=0$, i.e.\ if $j$ is divided by $p_i$. The number
of such zeroes in the range $0\ldots N-1$ is given by $N - \phi(N)$, where Euler's 
totient function $\phi(N)$ is
defined as the number of positive integers smaller than $N$ that are coprime to $N$.
In our case we have
\begin{equation}
  \label{eq:totient}
  \phi(N) = \prod_{i=1}^s p_i^{r_i-1}(p_i-1).
\end{equation}
A {\em Jacobi sequence} is a $\pm 1$ sequence of length $N$, given by
\begin{equation}
  \label{eq:def_jacobi_seq}
  s_j = \left\{
    \begin{array}{ll}
      \psi_N(j) & \text{if } \psi_N(j) \neq 0 \\
      \pm 1 & \text{if } \psi_N(j) = 0
    \end{array}
  \right..
\end{equation}
Setting the zeroes of the Jacobi symbol to $\pm 1$ leads to $2^{N-\phi(N)}$
distinct sequences. 

Jacobi sequences are potential candidates for low energy configurations, for three reasons:
First, the Jacobi sequences reduce to the Legendre sequences for $N$ prime.
Second, for $N=p_1p_2$  and a special choice for the zeroes of the Jacobi symbol, we
rediscover the two-prime sequences, eq.\ \myref{eq:two_prime_seq}. 
Third, Borsari et al \cite{borsari:96} have shown, that the
Jacobi sequences are ground states of the sine model, eq.\ \myref{eq:def_sine_model}, 
for $N=4t+3$ and not divisible by a square.

\begin{table}[htb]
\begin{center}
\begin{tabular}{r|rr|rrr}
 & & & \multicolumn{3}{c}{Energies} \\
N & p & q & two-prime & Jacobi & ground state \\\hline
21 & 7 & 3 & 84 & 84 & 42 \\
33 & 11 & 3 & 228 & 208 & 64 \\
39 & 13 & 3 & 774 & 342 & 86 \\
51 & 17 & 3 & 2306 & 610 & $\leq$ 146 \\
55 & 11 & 5 & 230 & 230 & $\leq$ 214\\
57 & 19 & 3 & 3672 & 648 & $\leq$ 184\\
65 & 13 & 5 & 736 & 592 & $\leq$ 224\\
69 & 23 & 3 & 7300 & 964 & $\leq$ 244\\
77 & 11 & 7 & 364 & 364 & $\leq$ 364\\
85 & 17 & 5 & 2100 & 1076 & $\leq$ 324\\
87 & 29 & 3 & 16118 & 1718 & $\leq$ 438\\
91 & 13 & 7 & 330 & 330 & $\leq$ 330\\
93 & 31 & 3 & 20508 & 1788 & $\leq$ 764\\
95 & 19 & 5 & 2926 & 1102 & $\leq$ 894
\end{tabular}
\end{center}
\caption[Tab1]{
  \label{tab:jacobi}
  Energies of configurations for $N=p*q$, $p,q$ odd primes and $p > q+2$. For $p=q+2$, twin prime
  sequences are ground states. The column ``Jacobi'' contains the minimum enery of all Jacobi sequences
  for given $N$, obtained by exhaustive enumeration of all $2^{N-\phi(N)}=2^{p+q-1}$ such sequences.
}
\end{table}

To check the suitability of the Jacobi sequences as low energy configurations for the Bernasconi model,
we did an exhaustive enumeration of all $2^{N-\phi(N)}$ Jacobi sequences of fixed length $N=pq$ to find
the sequence with minimum energy. Table \ref{tab:jacobi} displays the result. The two-prime sequences are
special Jacobi sequences, so their energy can not be lower than the Jacobi energy. The data from table
\ref{tab:jacobi} indicates, that the two-prime sequences have minimum energy among all Jacobi
sequences for
$p-q\leq 6$. For larger values of $p-q$, the energy of the two-prime sequences is larger. On the other
hand, the discrepancy between the Jacobi energy and the ground state energy is clearly visible. Except for
the twin-prime case, Jacobi sequences are not ground states of the Bernasconi model. This discrepancy is
even larger for non-squarefree values of $N$ (Table \ref{tab:jacobi_non_squarefree}).

\begin{table}[htb]
\begin{center}
\begin{tabular}{r|rr}
N & Jacobi & ground state \\\hline
9 & 56 & 24 \\
25 & 3000 & 72 \\
27 & 3802 & 74 \\
45 & 1372 & $\leq$ 124 \\
49 & 2055 & $\leq$ 144 \\
63 & 2622 & 62 \\
75 & 17354 & $\leq$ 298 
\end{tabular}
\end{center}
\caption[Tab1]{
  \label{tab:jacobi_non_squarefree}
  Minimum energies of Jacobi sequences for odd, non-squarefree values of $N<100$, compared to the
  ground state energies of the Bernasconi model.
}
\end{table}

This is no surprise, since prime powers dividing $N$ lead to regularities in the 
Jacobi symbol which keep the correlation values high. As an example consider $N=p^r$: 
In this case $\psi$ is periodic, $\psi_N(j+p) = \psi_N(j)$, and we get
\begin{equation}
  \label{eq:jacobi_estimate}
  C_k \geq p^{r-1}(p-2) \quad \text{ if }k\equiv 0 \bmod p
\end{equation}
for all Jacobi sequences.

%% file: conclusions.tex
\section{Summary and open problems}
\label{sec:conclusions}

We have demonstrated, that the ground states of the Bernasconi model with periodic
boundary conditions are closely related to cyclic or almost cyclic difference sets. 
Using theorems and well founded conjectures from the theory of cyclic
difference sets together with exact enumerations and numerical searches, we get
some results on the ground states of the Bernasconi model, which 
are conveniently summarized by 
the following facts and conjectures.
\begin{Fact}
\label{fact:perfect}
Perfect binary sequences, Eq.~\myref{eq:C_perfect} do exist and form ground states of the
Bernasconi model for $N=4,5,6,13$, $N=2^j-1$ with $j>1$,
$N=4t+3$ prime, $N=p(p+2)$ with both $p$ and $p+2$ prime.
\end{Fact}
This fact can be proven using the known constructions for cyclic difference sets,
as has been shown in section \ref{sec:perfect_sequences}. 
\begin{Con}
\label{con:perfect}
For all values of $N$ that do not match any of the conditions in Fact
\ref{fact:perfect}, there is no perfect sequence, i.e. ground states 
are formed by non-perfect sequences.
\end{Con}
This conjecture has a very solid base in the various non-existence theorems
for cyclic difference sets as cited in section \ref{sec:perfect_sequences}.
\begin{Con}
\label{con:almost}
Almost perfect sequences exist for all values of $N$.
\end{Con}
Beside those values of $N$ that match a condition in Fact \ref{fact:perfect},
this can be proven for $N=4t+1$ prime (Legendre sequence) and $N=q-1$, where
$q$ is an odd prime power (construction of Lempel et al.). For other values
of $N$, we checked this conjecture numerically up to $N<84$.
\begin{Con}
\label{con:almost_2}
For $N\equiv2\bmod4$, the ground states of the Bernasconi model are given
by almost perfect sequences. The ground state energy is $E=4(N-1)$.
For other values of $N$, the ground state energy obeys
\begin{equation}
  \label{eq:upper_limit}
  E \leq \left\{
  \begin{array}{ll}
    16(N-1) & \text{for $N\equiv0\bmod4$} \\
     9(N-1) & \text{for $N\equiv3\bmod4$}
  \end{array}
  \right.
\end{equation}
\end{Con}
This follows immediately from conjecture \ref{con:almost}.
\begin{Con}
\label{con:almost_1}
For $N\equiv1\bmod4$, almost perfect sequences are ground states of the
Bernasconi model. The ground state energy is $E=5(N-1)-8u(u+1)$, where
$u$ is a non-negative integer $\leq\frac12\sqrt{2N-1}$. 
\end{Con}
This conjecture is based on the numerical results from Table \ref{tab:almost_1}.
For the thermodynamic limit, our results lead us to
\begin{Con}
\label{con:limit}
The ground state energy of the Bernasconi model is $O(N)$. The limit
$\lim_{N\to\infty}\frac{E(N)}{N}$ is however not well defined, since the
precise value of $E(N)$ depends on number theoretic properties of $N$.
\end{Con}
This result is very different from the situation for the Bernasconi model with
aperiodic boundary conditions, where $E(N)=O(N^2)$, and $\lim_{N\to\infty}\frac{E(N)}{N^2}$
seems to be well defined \cite{mertens:96b}.

Marinari et al.~\cite{marinari:parisi:ritort:94a} noticed some patterns in their
numerical results for the ground state energies for $N\leq50$. Their equation~(12)
is a special case of conjecture \ref{con:almost}. Eq.~(13) of 
ref.~\cite{marinari:parisi:ritort:94a} is equivalent to conjecture \ref{con:almost_1}
with fixed value $u=3$ for $N\geq33$. Our numerical results (Table~\ref{tab:almost_1})
support this special value of $u$ up to $N=73$. For $N=77$ and $N=88$ we only found
$u\leq1$.

We did not find sequences with lower energies than those reported in 
ref.~\cite{marinari:parisi:ritort:94a}. This is due to the fact that numerical
methods like the one applied by Marinari et al.\ work quite well
if $N$ is not too large ($N \leq 50$). Our experience with numerical search suggests,
that beyond $N>100$ true ground-states can only be found by mathematical insight rather
than computer power.

One step towards more mathematical insight is to prove or disprove the above
conjectures. Especially the conjecture on the existence of almost perfect sequences for
all values of $N$ deserves some attention. Another open problem is the generalization of
successful construction methods for special $N$ to more general values. Here, the method
of Lempel et al.~\cite{lempel:cohn:eastman:77} is the most interesting candidate.

{\bf Acknowledgement:} The authors appreciate fruitful discussions with Alexander Pott.